# Real Options for Project Schedules (ROPS)

Lester Ingber

Lester Ingber Research
Ashland Oregon
ingber@ingber.com, ingber@alumni.caltech.edu
http://www.ingber.com/

**Abstract**

Real Options for Project Schedules (ROPS) has three recursive sampling/optimization shells. An outer Adaptive Simulated Annealing (ASA) optimization shell optimizes parameters of strategic Plans containing multiple Projects containing ordered Tasks. A middle shell samples probability distributions of durations of Tasks. An inner shell samples probability distributions of costs of Tasks. PATHTREE is used to develop options on schedules. Algorithms used for Trading in Risk Dimensions (TRD) are applied to develop a relative risk analysis among projects.

---

## 1. Introduction

This paper is a brief description of a methodology of developing options (in the sense of financial options, e.g., with all Greeks), to be applied in collaboration with Michael Bowman, as a first example to scheduling a massive US Army project, Future Combat Systems (FCS) [1].

The major focus is to develop Real Options for non-financial projects, as discussed in other earlier papers [3,4,12]. Data and some guidance on its use has been reported in a previous study of FCS [2,5]. The need for tools for fairly scheduling and pricing such a complex project has been emphasized in Recommendations for Executive Action in a report by the U.S. General Accounting Office (GAO) on FCS [14], and they also emphasize the need for management of FCS business plans [13].

## 2. Goals

A given Plan results in $S(t)$, money allocated by the client/government is defined in terms of Projects $S_i(t)$,

$$S(t) = \sum_i S_i(t)$$

where $a_i(t)$ may be some scheduled constraints. PATHTREE processes a probability tree developed over the life of the plan $T$, divided into $N$ nodes at times $\{t_n\}$, each with mean epoch length $dt$ [11]. Options, including all Greeks, familiar to financial markets, are calculated for quite arbitrary nonlinear means and variances of multiplicative noise [6,9]. This ability to process nonlinear functions in probability distributions is essential for real-world applications.

Each Task has a range of durations, with nonzero $A_i$, with a disbursement of funds used, defining $S_i(t_n)$. Any Task dependent on a Task completion is slaved to its precursor(s).

We develop the Plan conditional probability density (CPD) in terms of differenced costs, $dS$,

$$P(S \pm dS; t_n + dt | S; t_n)$$

P is modeled/cast/fit into the functional form

$$P(S \pm dS; t_n + dt | S; t_n) = (2\pi g^2 dt)^{-\frac{1}{2}} \exp(-L dt)$$

$$L = \frac{(dS - f dt)^2}{(2 g^2 dt^2)}$$

where $f$ and $g$ are nonlinear function of cost $S$ and time $t$. The $g^2$ variance function absorbs the multiple Task cost and schedule statistical spreads, to determine $P(dS, t)$, giving rise to the stochastic nature of dollars spent on the Plan.

A given Project $i$ with Task $k$ has a mean duration $i_{ik}$, with a a mean cost $S_{ik}$. The spread in $dS$ has two components arising from: (1) a stochastic duration around the mean duration, and (2) a stochastic spread of mean dollars around a deterministic disbursement at a given time. Different finite-width asymmetric distributions are used for durations and costs. For example, the distribution created for Adaptive Simulated Annealing (ASA) [8], originally called Very Fast Simulated Re-annealing [7], is a finite-ranged distribution with shape determined by a parameter "temperature" $q$. For each state (whether duration or cost): (a) A random binary choice can be made to be higher or lower than the mean, using any ratio of probabilities selected by the client. (b) Then, an ASA distribution is used on the chosen side. Each side has a different $q$, each falling off from the mean. This is illustrated and further described in Fig. 1.

At the end of the tree at a time $T$ ($T$ also can be a parameter), there is a total cost at each node $S(T)$, called a final "strike" in financial language. (A final strike might also appear at any node before T due to cancellation of the Project using a particular kind of schedule alternative.) Working backwards, options are calculated at time $t_0$. Greeks (functional derivatives of the option) assess sensitivity to various variables, e.g., like those discussed in previous papers [12], but here we deliver precise numbers based on as much real-world information available.

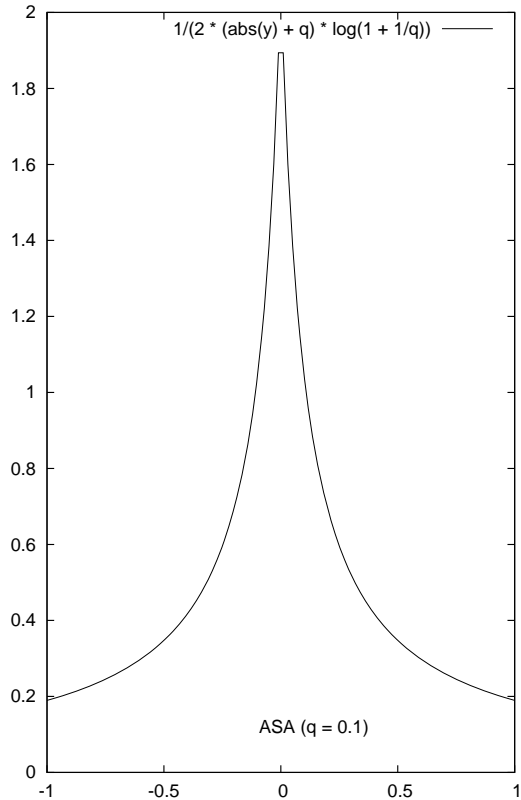

Fig. 1. The ASA distribution can be used to develop finite-range asymmetric distributions from which a value can be chosen for a given state of duration or cost. (a) A random binary distribution is selected for a lower-than or higher-than mean, using any ratio of probabilities selected by the client. Each side of the mean has its own temperature $q$. Here an ASA distribution is given for $q = 0.1$. The range can be scaled to any finite interval and the mean placed within this range. (b) A uniform random distribution selects a value from [-1,1], and a normalized ASA value is read off for the given state.

## 3. Data

The following data are used to develop Plan CPD. Each Task $i$ has
   (a) a Projected allocated cost, $C_i$
   (b) a Projected time schedule, $T_i$
   (c) a CPD with a statistical width of funds spent, $SW_{S_i}$
   (d) a distribution with a statistical width of duration, $SW_{T_i}$
   (e) a range of durations, $R_{T_i}$
   (f) a range of costs, $R_{S_i}$
Expert guesses need to be provided for (c)-(f) for the prototype study.

A given Plan must be constructed among all Tasks, specified the ordering of Tasks, e.g., obeying any sequential constraints among Tasks.

## 4. Three Recursive Shell

### 4.1. Outer Shell

There may be several parameters in the Project, e.g., as coefficients of variables in means and variances of different CPD. These are optimized in an outer shell using ASA [8]. This end product, including MULTI_MIN states returned by ASA, gives the client flexibility to apply during a full Project [12]. We may wish to minimize Cost/$T$, or (CostOverrun - CostInitial)/$T$, etc.

### 4.2. Middle Shell

To obtain the Plan CPD, an middle shell of Monte Carlo (MC) states are generated from recursive calculations. A Weibull or some other asymmetric finite distribution might be used for Task durations. For a given state in the outer middle, a MC state has durations and mean cost disbursements defined for each Task.

### 4.3. Inner Shell

At each time, for each Task, the differenced cost $(S_{ik}(t+dt) - S_{ik}(t))$ is subjected to a inner shell stochastic variation, e.g., some asymmetric finite distribution. The net costs $dS_{ik}(t)$ for each Project $i$ and Task $k$ are added to define $dS(t)$ for the Plan. The inner shell cost CPD is re-applied many times to get a set of $\{dS\}$ at each time.

## 5. Real Options

### 5.1. Plan Options

After the Outer MC sampling is completed, there are histograms generated of the Plan's $dS(t)$ and $dS(t)/S(t-dt)$ at each time $t$. The histograms are normalized at each time to give $P(dS, t)$. At each time $t$, the data representing $P$ is "curve-fit" to the form of Eq. (0), where $f$ and $g$ are functions needed to get good fits, e.g., fitting coefficients of parameters $\{x\}$

$$f = x_{f0} + x_{f1}S + x_{f2}S^2 + \cdots$$

$$g = x_{g0} + x_{g1}S + x_{g2}S^2 + \cdots$$

At each time $t$, the functions $f$ and $g$ are fit to the function $\ln(P(dS,t))$, which includes the prefactor containing $g$ and the function $L$ which may be viewed as a Padé approximate of these polynomials.

Complex constraints as functions of $S_{ik}(t)$ can be easily incorporated in this approach, e.g., due to regular reviews by funding agencies or executives. These $P$'s are input into PATHTREE to calculate options for a given strategy or Plan.

### 5.2. Risk Management of Project Options

If some measure of risk among Projects is desired, then during the MC calculations developed for the top-level Plan, sets of differenced costs for each Project, $dS_i(t)$ and $dS_i(t)/S_i(t-dt)$, stored from each of the Project's Tasks. Then, histograms and Project CPDs are developed, similar to the development of the Plan CPD. A copula analysis, coded in TRD for risk management of financial markets, are applied to develop a relative risk analysis among these projects [10]. In such an analysis, the Project marginal CPDs are all transformed to Gaussian spaces, where it makes sense to calculate covariances and correlations. An audit trail back the original Project spaces permits analysis of risk dependent on the tails of the Project CPDs.

## 6. Generic Applications

ROPS can be applied to any complex scheduling of tasks similar to the FCS project. The need for government agencies to plan and monitor such large projects is becoming increasingly difficult and necessary [15]. Many large businesses have similar projects and similar requirements to manage their complex projects.